# Cascaded 3D Full-body Pose Regression from Single Depth Image at 100 FPS

Shihong Xia*
Institute of Computing Technology, CAS

Zihao Zhang†
Institute of Computing Technology, CAS
University of Chinese Academy of Sciences

Le Su‡
Civil Aviation University of China

## ABSTRACT

There are increasing real-time live applications in virtual reality, where it plays an important role in capturing and retargetting 3D human pose. But it is still challenging to estimate accurate 3D pose from consumer imaging devices such as depth camera. This paper presents a novel cascaded 3D full-body pose regression method to estimate accurate pose from a single depth image at 100 fps. The key idea is to train cascaded regressors based on Gradient Boosting algorithm from pre-recorded human motion capture database. By incorporating hierarchical kinematics model of human pose into the learning procedure, we can directly estimate accurate 3D joint angles instead of joint positions. The biggest advantage of this model is that the bone length can be preserved during the whole 3D pose estimation procedure, which leads to more effective features and higher pose estimation accuracy. Our method can be used as an initialization procedure when combining with tracking methods. We demonstrate the power of our method on a wide range of synthesized human motion data from CMU mocap database, Human3.6M dataset and real human movements data captured in real time. In our comparison against previous 3D pose estimation methods and commercial system such as Kinect 2017, we achieve the state-of-the-art accuracy.

## 1 INTRODUCTION

Motion capture technology has been widely used in virtual reality and entertainment industry. With the development of consumer devices such as depth camera, accurately estimating human pose has been the fundamental and key problem, especially from only one camera and many efforts have been made to capture human motion from color or depth image sequences. As stated in [25], recent works on image-based human pose capture can be divided into two methods, namely, color image based and depth image based method. However, both methods have the pose ambiguity problem. Compared with RGB image based methods, the extra spatial information provided by utilizing the depth image for 3D skeletons extraction could yield more practical solutions. Therefore, we still focus on the depth image based capture methods.

There exists generative and discriminative methods in depth image based motion capture community. The generative methods [3, 19], also known as model-based methods, need to build the prior knowledge of human shape model and include modeling and estimation steps. The discriminative methods, also known as model-free methods, directly estimate human pose from single depth image without the prior knowledge of human shape model. The main challenges of model-free methods are the self-occlusion and the information loss caused by monocular camera, which makes it hard to get accurate 3D pose of full human body. So, how to accurately estimate the joint with depth image becomes a hot topic in recent years. Many solutions have been proposed with the help of decision forest. [13, 21, 26]. However, each decision forest only corresponds to one regression/classification model, which means it might be diffcult to specify the complex relationship between the depth information and the joint position.

*e-mail: xsh@ict.ac.cn
†e-mail: zhangzihao@ict.ac.cn
‡e-mail:sule@ict.ac.cn

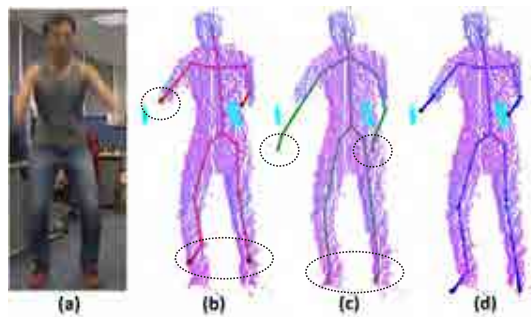

Figure 1: The accuracy example of our method. (a) The RGB image captured by Kinect; (b) The estimated pose of cascaded based method without kinematic model; (c) The original estimated pose by Kinect; (d) The result of our method. The inaccurate estimation made by state-of-the-art methods are highlighted by circles.

Inspired by the cascaded hand pose regression method [22] and feature points regression of human face [6], in this paper, we assume that the more regression/classification models we have, the more accurate human pose we can get. This assumption leads us to investigate a cascaded pose regression model which includes more than one regression models during the pose estimation step. We first evaluate the off-the-shelf product in this community, by incorporating the well-known decision forest methods by Kinect. Also we investigate the traditional cascaded methods. Figure 1 illustrates test results on real motion capture data and reveals the limitation of previous methods. What we present here is a novel cascaded 3D full-body pose regression method to estimate accurate pose from a single depth image at 100 fps. The key idea is to train cascaded regressors based on Gradient Boosting algorithm from pre-recorded human motion capture database. Moreover, we further incorporate hierarchical kinematics model of human pose into the learning procedure so that we can directly estimate accurate 3D joint angles instead of joint positions. In our comparison against previous 3D pose estimation methods and commercial system such as Kinect 2017, we achieve the state-of-the-art accuracy, probably because traditional methods do not consider the kinematic constraints such as the bone length or the joint angle.

Our method has the following technical contributions:

- A novel general framework for accurate 3D full-body pose estimation from a single depth image based on Gradient Boosting algorithm for learning an ensemble of regression forests.

- A novel 3D full-body pose regression model which integrates

kinematics model and enables more effective features and more accurate estimation.
- An accurate 3D full-body pose estimation system at 100fps, which is capable of estimating heterogeneous motion.

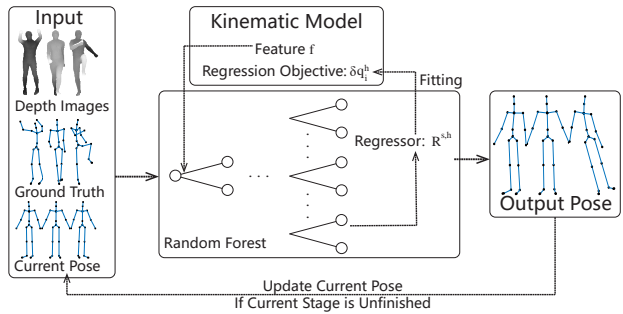

Figure 2: Our algorithm overview: the algorithm includes three training stages from which we can get the regressor of certain joint. We only show the torso stage and its results in this figure.

## 2 RELATED WORK

In this section, we focus on previous works on 3D human pose estimation from a single depth image, which are most relevant to the context of this paper. We will also discuss related works on cascaded regression methods and imposing kinematic model for pose estimation.

### 2.1 3D Pose Estimation from Depth Images

The skeletal pose estimation from images and videos is an important task, but still has many challenges. Different approaches based on machine learning or statistic methods have been proposed [13, 15, 21, 24] to solve this problem. Many human pose estimation methods tend to use decision forests as the regressors. The state-of-the-art work [21], adopts the random forest classifier to do the pixel-level classification. Then they employ cluster algorithm to obtain the body part information. But this method will fail in facing with occlusion. Girshick and his colleagues [13] further extend this work by using Hough forests to cast per-pixel vote for the joint position, and can get better results than Shotton [21], especially in occlusion case. The work [24] also uses regression forest to obtain the correspondence between body surface and the pixels. This work shows its robustness by fitting to different body size. However, it may meet the problems of miscorrespondence, and heavily relies on the training samples. In the work of [15], the authors adopt the geodesic distance as the feature to avoid the ambiguities in pose estimation, but it mainly aims at estimating the upper body pose. Recently, in [18], the authors optimize the method [13] by introducing a random verification forest so that the vote can be much more accurate. However, they only test their methods on the homogeneous motion. In this paper, we also use the random decision forests as regressors, but differently, we build it in a cascaded manner so that we can get more than one regression model and can ensure a better description of the complex relationship between the depth information and the human pose. The experiments turn out that we can achieve more accurate results compared with those state-of-the-art methods [13, 21].

### 2.2 Cascaded Regression Method

Our approach is directly motivated by the cascade regression works [9, 20, 22]. They adopt cascaded regression method to fulfill different regression tasks, such as face detection [9] and hand pose regression. In the work of [20], the authors learn a coarse-to-fine cascaded model to estimate the 2D human pose in images. In the work of [22], the authors apply cascaded regression method to solve the hand pose estimation problems. They also claim that this method can also be applied to human body pose estimation problems, but still it lacks the consideration of kinematic constraints. Our method, instead, uses a kinematic model during the regression process to ensure our approach is capable of modeling the complexity of different human poses. This kinematic model leads to more effective features and more accurate pose estimation.

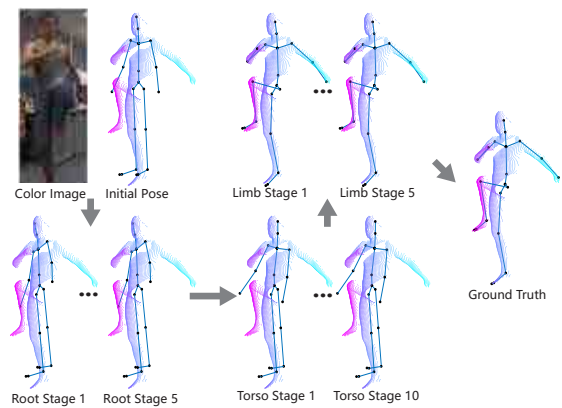

Figure 3: An real case demonstration of our method.

### 2.3 Kinematic Model Based Method

There are also some works [5,8,27] taking the kinematic information into consideration. In the work of [5], the authors propose an open-source system which can estimate human pose without any pre-process such as background subtraction or environment mapping. The kinematic model applied in this work is mainly limited in the distance between joints. However, the kinematic information of the ground truth and the correspondence between the depth information and skeleton are not taken into consideration. The work by Zhou and his colleagues [27] uses the forward kinematic method as a special layer in the networks so that they can get the joint loss from their motion parameters. Different from their works, which model the bone length or joint angle, we directly use a higher level kinematic parameters, the pose gradient, as the regression objective. We also extract the random forests feature based on the kinematic chain. In the work of [8], the authors propose a generalized Gaussian kernel correlation embed with a kinematic skeleton for the articulated pose estimation. Their method is based on a Gaussian kernel model, while our method is totally model-free.

## 3 METHOD OVERVIEW

The algorithm overview and a real case demonstration are shown in Figure 2 and Figure 3, respectively. We employ the hierarchy method in the whole training process which includes three different stages: the root stage, the torso stage and the limb stage. As shown in Figure 2, we adopt a novel kinematic model including the kinematic related features and the negative gradient residual between the ground truth and the current pose as our regression objective. Each training sample corresponds to an output pose after a training process, and the output is then used to update our current pose if current training stage is unfinished.

As for the testing process, given a set of single captured depth images and rough 3D initial full-body poses (see Section 4.3), the root position and orientation are estimated via 5-stage cascaded regressor at first (see the left bottom part of Figure 3). Then, torso joints' relative rotations are estimated through 10-stage cascaded regressor based on root estimation result (see the middle bottom part of Figure 3). Finally, four limbs joints' relative rotations are estimated by 5-stage cascaded regressors based on torso estimation result, separately (see the upper middle part of Figure 3). In our proposed cascaded regression model, we use random forests [4] as the weak learner at each regression stage.

**Depth data acquisition.** Current commercial depth cameras are low-cost, easy to deploy and can record 3D depth data at a relatively high frame rate. In this paper, we use Microsoft Kinect [2] cameras, which cost roughly a hundred dollars, giving a 512×424 pixels depth image at 30 frames per second (fps) with depth resolution of a few centimeters. Pixels in a depth image $\mathbf{I}$ store calibrated depth data in the scene, rather than a measure of intensity or color. In our experiment, each pixel $\mathbf{x} = (u,v)^T \in R^2$ in the depth image $\mathbf{I}$ stores the depth value $d(\mathbf{x})$ and its corresponding x-y-z coordinates $\mathbf{p} = (x,y,z)^T \in R^3$ in the 3D space. We use the Kinect API to extract foreground pixels, and keep their depth values unchanged while assigning depth values of background pixels to 1e5.

**Full-body pose representation.** We apply full-body skeleton model by using a hierarchical tree structure which contains 16 rigid body segments, including waist, lower back, upper back, head, and left/right clavicle, upper arm, lower arm, upper leg, lower leg, foot, and toes. We describe the 3D full-body pose by using a set of independent joint coordinates $\mathbf{q} \in R^{38}$, including absolute root position and orientation as well as the relative joint angles of individual joints. These joints are the root (6 Dof), head (3 Dof), upper back (3 Dof), left and right clavicle (2 Dof), humerus (3 Dof), radius (1 Dof),. femur (3 Dof), tibia (1 Dof), and foot (3 Dof). The notion "Dof" is short for "Degree of freedom"

**Human pose database.** For the purpose of our cascaded regression model training and testing, we construct a heterogeneous 3D full-body pose database from the synthetic data based on CMU mocap database [1]. Our constructed pose database contains approximate 1.5 hours, including: walking, running, boxing, kicking, jumping, dancing, marching, gymnastics, golf and hand waving. We use an existing efficient motion retargeting technique [14] to accomplish the skeleton normalization tasks.

**Automatic subject calibration.** Since the prior knowledge of body size (length and radius of each body segment) is necessary during the training and testing procedure in our proposed cascaded regression method, we introduce a simple classification forest based learning method to address this problem inspired by the work [21]. The key idea is to train a body-part classification forest on prerecorded T-pose RGB-D images to estimate new users' body parts. Firstly, we build a depth dataset which only includes the recorded T-poses from various human actors. Secondly, we learn a pixel-level body-part label distribution. Then we define a density estimator per body label based on the body part distribution and use the mean shift algorithm to find the joint position. Since each new user only has to perform this procedure once and this method is based on classification forest, it is practicable, accurate and fast enough for real-time live applications.

## 4 CASCADED 3D FULL-BODY POSE REGRESSION

Given a captured depth image $\mathbf{I}$, a rough initial 3D full-body pose $\mathbf{q}^0$ and trained cascaded regressors $\{\mathbf{R}^h\}_{h=1}^H$, we can get the result 3D full-body pose $\widetilde{\mathbf{q}} = \mathbf{q}^H$ via an iterative stage-wise regression procedure $(h = 1, \cdots, H)$. According to method [22], the common 3D full-body pose update within each stage is as follows:

$$\mathbf{q}^h = \mathbf{q}^{h-1} + R^h(\mathbf{I}, \mathbf{q}^{h-1}) \quad (1)$$

where the final pose estimation result is $\mathbf{q}^H$. During the offline cascaded 3D full-body pose regression training period, we propose the hierarchical regression method along by the tree structural human skeleton (see Section 4.1). For cascaded regression model training of each body parts, we define the 3D pose gradient $\delta \mathbf{q}^h$ as regression target based on Gradient Boosting algorithm (the definitions can be seen in Section 4.2.1). In each stage $h$, we train a random forest as the 3D pose regressor $\mathbf{R}^h$ by approximating the defined regression target $\delta \mathbf{q}^h$; to train the weak learner on the splitting node of decision tree of random forests at stage $h$, we define kinematics chain based 3D pose-indexed features and adaptive 3D features sampling method (see Section 4.2.2). At each stage $h$, we can significantly increase the probability of getting useful features by incorporating kinematics chain into the 3D pose-indexed features sampling procedure according to 3D pose update from last stage $h - 1$ and current captured depth image.

### 4.1 Hierarchical Regression Strategy

For highly articulated structure as human full-body, we propose hierarchical regression strategy by making the following key observations:

1. The global position and orientation of 3D full-body pose mainly rely on root joint.

2. The pose variations between torso and limb are significantly different. Torso has less pose variance than limb. Torso and limb pose regressions should be done separately as well as four limbs pose regressions to improve the accuracy and speed up the convergence of boosted regression framework [11], because each weak regressor has relatively large errors.

3. The articulated structure of 3D full-body pose indicates that root pose severely affects torso pose, and torso pose severely affects limbs poses. More specifically, a large number of variations in the torso pose are caused by different root pose, and a large number of variations in the limbs poses are caused by the changes in the torso pose, rather than torso and limbs articulations. Therefore, in order to improve full-body pose regression accuracy, after regressing root pose, we refine it during torso regression process and refine torso pose during limbs regression process.

We captured massive human movements from actual life and made all the above observations, which have strong objective basis. Therefore, we propose hierarchical regression method for human full-body pose estimation, as illustrated in Figure 3. Firstly, we only regress the global position and orientation of root joint. Secondly, we regress the torso joints' pose based on root pose. Finally, we regress the limbs joints' poses separately based on torso joints' poses. According to our 3D full-body pose representation in Section 3, torso joints include: upper neck, r/lclavicle, r/lhumerus, neck, head and r/lfemur, and limbs joints include: r/lhumerus, r/lradius, r/lfemur, r/ltibia and r/lfoot. For the purpose of improving pose estimation accuracy, we suggest that when regressing limbs' poses, joints r/lhumerus and r/lfemur need considering in that regression process as well.

### 4.2 Kinematic Regression Model

For the highly complex human pose such as kicking or boxing, due to the self-occlusion, carrying out the regression on the joint position may not be enough for the weak learner to model these human poses. So, we propose a kinematic model which contains two parts, the kinematic regression objective and the kinematic chain based feature. In this section, we will describe them separately.

#### 4.2.1 Gradient Boosting based Regression

Based on Gradient Boosting algorithm [11, 12, 17], we first define the Loss Function of our proposed cascaded 3D full-body pose regression as follows:

$$L = \arg\min_{\mathbf{q}_i} \sum_{i=1} \|J(\mathbf{q}_i) - J(\mathbf{q}_i^*)\|^2 \quad (2)$$

where $J(\mathbf{q}_i)$ and $J(\mathbf{q}_i^*)$ represent 3D joints' positions of estimated pose $\mathbf{q}_i$ and its corresponding ground truth pose $\mathbf{q}_i^*$. We use forward kinematic technique to calculate them based on prior knowledge of current user's skeleton model. $i$ represents $i^{th}$ training sample.

Then, we define 3D pose regression target at stage $h$ as the negative gradient of loss function $L$ (Eq. 2):

$$\delta \mathbf{q}_i^h = -\frac{\partial J(\mathbf{q}_i^{h-1})}{\partial \mathbf{q}_i^{h-1}} * (J(\mathbf{q}_i^{h-1}) - J(\mathbf{q}_i^*)) \qquad (3)$$

where $J(\mathbf{q}_i^{h-1})$ and $J(\mathbf{q}_i^*)$ represent 3D joints' positions of 3D pose $\mathbf{q}_i^{h-1}$ and its corresponding ground truth pose $\mathbf{q}_i^*$ at stage $h-1$ of training sample $i$. $\frac{\partial J(\mathbf{q}_i^{h-1})}{\partial \mathbf{q}_i^{h-1}}$ represents Jacobian Matrix of 3D joints' positions $J(\mathbf{q}_i^{h-1})$ with respect to 3D pose $\mathbf{q}_i^{h-1}$ at stage $h-1$. In our experiments (see Section 5), we will show the outperformed accuracy of our proposed negative 3D pose gradient as regression target comparing with directly joints' 3D positions [22] and Euler angle which is most commonly used for 3D human pose representation in order to support that our regression objective is the best.

**3D pose update.** According to Gradient Boosting algorithm [11, 12, 17], the learnt 3D pose regressor $R^h$ at stage $h$ is only the approximate direction of the negative gradient $g^h = R^h(\mathbf{I}, \mathbf{q}_i^{h-1})$ of the loss function $L$ (Eq. 2). Then we calculate the optimal step size of that direction by minimizing the following objective function:

$$\widetilde{\beta}^h = \arg\min_{\beta^h} \sum_{i=1} \|J(\mathbf{q}_i^{h-1} + \beta^h \cdot g^h) - J(\mathbf{q}_i^*)\|^2 \qquad (4)$$

where the step size $\beta^h$ at stage $h$ is a scalar variable. In this paper, we calculate $\beta^h$ by the Line Search algorithm [23] to Eq. 4. Then, we update the 3D pose $\mathbf{q}_i^h$ at current stage $h$ as follows:

$$\mathbf{q}_i^h = \mathbf{q}_i^{h-1} + \widetilde{\beta}^h \cdot g^h \qquad (5)$$

where $\mathbf{q}_i^{h-1}$ is 3D pose result at stage $h-1$, $g^h$ and $\widetilde{\beta}^h$ are the approximate negative gradient and its step size of loss function at stage $h$, respectively.

### 4.2.2 Kinematics based 3D Pose-indexed Features

Similar to existing learning based human pose estimation from depth image methods [21], we also use depth pixel difference, such as $f = d(\mathbf{I}(\mathbf{x}+\mathbf{u}_1)) - d(\mathbf{I}(\mathbf{x}+\mathbf{u}_2))$, as 3D pose feature for training the splitting nodes of decision trees within random forests. $\mathbf{u}_1$ and $\mathbf{u}_2$ are two 2D pixel offsets. In order to acquire certain geometric invariance for $\mathbf{u}_i (i = 1,2)$, which is one of the most essential factors for our regression model, we define kinematics based 3D pose-indexed features as follows:

$$f = CamProj(\mathbf{T}_{\mathbf{q}^{h-1},i}(J_i(\mathbf{q}^c) + \Delta \mathbf{p}_1^c)) \qquad (6)$$
$$\quad - CamProj(\mathbf{T}_{\mathbf{q}^{h-1},j}(J_j(\mathbf{q}^c) + \Delta \mathbf{p}_2^c))$$

where $\mathbf{q}^{h-1}$ is 3D pose at stage $h-1$, $\mathbf{q}^c$ is 3D canonical pose, $J_i(\mathbf{q}^c)$ and $J_j(\mathbf{q}^c)$ are 3D joints' positions under 3D canonical pose, which are calculated by forward kinematic technique. $\Delta \mathbf{p}_1^c$ and $\Delta \mathbf{p}_2^c$ are 3D offsets under 3D canonical pose space. $\mathbf{T}_{\mathbf{q}^{h-1},i}$ and $\mathbf{T}_{\mathbf{q}^{h-1},j}$ are kinematic based 3D world transformation matrices, which transform 3D points $J_i(\mathbf{q}^c) + \Delta \mathbf{p}_1^c$ and $J_j(\mathbf{q}^c) + \Delta \mathbf{p}_2^c$ from 3D canonical pose $\mathbf{q}^c$ space to current 3D pose space at stage $h$ according to 3D pose $\mathbf{q}^{h-1}$ at stage $h-1$. We define $\mathbf{T}_{\mathbf{q}^{h-1},i}$ and $\mathbf{T}_{\mathbf{q}^{h-1},j}$ as follows:

$$\mathbf{T}_{\mathbf{q}^{h-1},i} = \mathbf{wR}(\mathbf{q}^{h-1},i) \cdot \mathbf{IT}(\Delta \mathbf{p}_1^c) \qquad (7)$$
$$\mathbf{T}_{\mathbf{q}^{h-1},j} = \mathbf{wR}(\mathbf{q}^{h-1},j) \cdot \mathbf{IT}(\Delta \mathbf{p}_2^c)$$

where $\mathbf{wR}(\mathbf{q}^{h-1},i)$ and $\mathbf{wR}(\mathbf{q}^{h-1},j)$ are the $i^{th}$ and $j^{th}$ joints' world rotation matrices (calculated by forward kinematics technique) of

3D pose at stage $h$. $\mathbf{IT}(\Delta \mathbf{p}_1^c)$ and $\mathbf{IT}(\Delta \mathbf{p}_2^c)$ are 3D offsets $\Delta \mathbf{p}_1^c$ and $\Delta \mathbf{p}_2^c$ with respect to 3D translation matrices of joint $i$ and joint $j$. In our experiment, 3D offsets $\Delta \mathbf{p}_1^c$ and $\Delta \mathbf{p}_2^c$ under the canonical pose $\mathbf{q}^c$ are randomly sampled within a predefined 3D spherical bounding box whose radius is related to length and radius of specific body parts; there is fifty-fifty chance for joint $i$ and $j$ to be the same (unary feature) or not (binary feature). Our kinematics 3D pose-indexed feature extraction procedure works as follows:

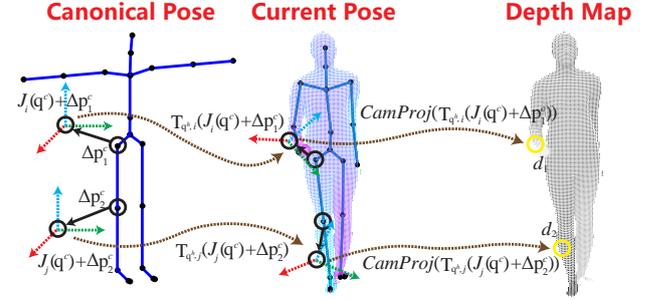

Figure 4: An illustration of kinematics 3D pose-indexed 2D feature extraction for right upper-leg regressor training.

1. Randomly generate two 3D offsets $\Delta \mathbf{p}_2^c$ and $\Delta \mathbf{p}_2^c$ within predefined 3D spherical bounding box for joint $i$ and $j$ at current stage $h$ under the 3D canonical pose $\mathbf{q}^c$ space.

2. Calculate 3D world positions $\mathbf{T}_{\mathbf{q}^{h-1},i}$ and $\mathbf{T}_{\mathbf{q}^{h-1},j}$ for two randomly sampled 3D offsets $\Delta \mathbf{p}_2^c$ and $\Delta \mathbf{p}_2^c$ under the 3D pose $\mathbf{q}^{h-1}$ space.

3. Project two 3D world positions $\mathbf{T}_{\mathbf{q}^{h-1},i}$ and $\mathbf{T}_{\mathbf{q}^{h-1},j}$ of 3D offsets $\Delta \mathbf{p}_2^c$ and $\Delta \mathbf{p}_2^c$ to 2D depth pixels on the camera image plane, then calculate pixels' depth difference as the feature $f$.

An example of kinematics 3D pose-indexed 2D feature extraction for right upper-leg regressor training is shown in Figure 4.

**Adaptive 3D Features Sampling** As introduced above, we employ the hierarchy method to estimate the 3D full-body pose. During the regression process, we follow the order of "Root Joint $r \rightarrow$ Torso Joints $t \rightarrow$ Limb Joints $l$", and the features are randomly extracted from the sphere bounding box of the joint center. Due to the difference in the bone length of each joint, we propose an adaptive 3D feature sampling method. That is to set the radius limbs 3D feature sampling radius $dr^l$ as the criterion radius, and the root joint and the torso joints' 3D feature sampling radius $dr^r$ and $dr^t$ change over the ratio with $dr^l$

$$dr^r = \frac{Bl^r \cdot Br^r}{Bl^l \cdot Br^l} \cdot dr^l, \quad dr^t = \frac{Bl^t \cdot Br^t}{Bl^l \cdot Br^l} \cdot dr^l \qquad (8)$$

The $Bl^r$ and $Br^r$ are the bone length of the root joint part and radius respectively. The $Bl^l$ and $Br^l$ are the bone length of the limb joint part and radius respectively.

### 4.3 Implementation Details

In this section, we will give the training and testing details of our proposed cascaded regression model, including: initial pose, offline training and online testing procedure.

**Initial Pose.** As mentioned in Section 4.2.1, we define 3D pose regression target as negative gradient of loss function for our proposed cascaded 3D full-body pose regression model. At the beginning of either training or testing process, we use the same 3D full-body pose as the initial pose. In our experiments, the initial pose is an "A"-pose, which is represented by $\mathbf{q}^0$. For each training and testing sample $i$, the initial world position of root joint is calculated as the mean position of the input 3D depth point cloud by Mean Shift algorithm [7].

**Algorithm 1** Offline Training of Hierarchical 3D Full-body Pose Regression

---
**Require:** depth map $\mathbf{I}_i$, ground truth pose $\mathbf{q}_i^*$, and initial pose $\mathbf{q}_i^0$ for all training samples $i$
**Ensure:** Different stages' regressor $\{R^{s,h}\}_{h=1}^{H_r}, s \in \{r,t,l\}$
1: **for** $s \in \{r,t,l\}$ **do**  ▷ Traverse all the stages
2:    **for** $h = 1 \to H_s$ **do**  ▷ learn current stage's regressor
3:      $\delta \mathbf{q}_i^{s,h} = -\frac{\partial J(\mathbf{q}_i^{s,h-1})}{\partial \mathbf{q}_i^{s,h-1}} * (J(\mathbf{q}_i^{s,h-1}) - J(\mathbf{q}_i^{s,*}))$  ▷ negative gradient residual of current stage
4:      learn $R^{s,h}$ to approximate $\delta \mathbf{q}_i^{s,h}$
5:      calculate approximated gradient $g^{s,h} = R^{s,h}(\mathbf{I}, \mathbf{q}_i^{s,h-1})$
6:      $\widetilde{\beta}^{r,h} = \arg\min_{\beta^{r,h}} \sum_{i=1} \|J(\mathbf{q}_i + \beta^{s,h} \cdot g^{s,h}) - J(\mathbf{q}_i^*)\|^2$  ▷ learn step size
7:      $\mathbf{q}_i^{s,h} = \mathbf{q}_i^{s,h-1} + \widetilde{\beta}^{s,h} \cdot g^{s,h}$  ▷ update current stage's pose
8:    **end for**
9:    Initialize $\mathbf{q}_i^{t,0}$ as canonical torso poses on $\mathbf{q}_i^{s,H_s}$
10:   $\mathbf{q}_i^0 = \mathbf{q}_i^{s,H_s} \cup \mathbf{q}_i^{t,0}$  ▷ re-initialize whole body pose
11: **end for**
12: $\mathbf{q}_i^h = \mathbf{q}_i^{t,H_t} \cup \{\mathbf{q}_i^{l,h}\}_{l \in limbs}$  ▷ update whole body pose

---

**Offline Training.** Input data are training dataset $(\mathbf{I}_i, \mathbf{q}_i^*), i = 1, \cdots, N$, where $\mathbf{I}$ and $\mathbf{q}_i^*$ are depth image and corresponding 3D ground truth full-body pose of training sample $i$, respectively. Based on our proposed hierarchical regression method, our 3D pose training order is as follows: "root pose $r \to$ torso pose $t \to$ limbs $l$", and the pseudo training algorithm of our approach is shown in Algorithm 1, where $H_r$, $H_t$ and $H_l$ are stage numbers of root regressor, torso regressor and limbs regressors, respectively. $\mathbf{q}_i^{r,h}$, $\mathbf{q}_i^{t,h}$ and $\mathbf{q}_i^{l,h}$ are 3D pose of root, torso and limbs at stage $h$.

**Online Testing.** Input data are testing depth image $I$ and 3D initial pose $\mathbf{q}^0$. The hierarchical testing order is as follows: "root pose $r \to$ torso $t \to$ limbs $l$"(as shown in Figure 3), and the pseudo testing algorithm of our approach is shown in Algorithm 2, where $\widetilde{\beta}^{r,h}$, $\widetilde{\beta}^{t,h}$ and $\widetilde{\beta}^{l,h}$ are learned step sizes of negative gradient of loss function at stage $h$ for root, torso and limbs, respectively.

## 5 EXPERIMENTS

In this section, we will firstly show effects on pose estimation accuracy by evaluating different choices of the hyper-parameters of our proposed cascaded regression model, including: tree depth, number of trees per stage, stage number and maximum 3D offset sampling radius. Then, we will discuss three different kinds of regression objectives including 3D joint positions, joint rotation angle (Euler angle) and 3D pose changing gradient. Thirdly, we will show the outperformed pose estimation accuracy by comparing with several alternative state-of-the-art methods [13, 21, 22]. Finally, we will show the outperformed pose estimation efficiency by comparing with several alternative state-of-the-art methods [13, 21]. We use several RGB-D datasets for the evaluations, including synthesized depth images according to CMU mocap database, depth images from Human3.6M [10, 16], and real-time live captured depth images by ourself. All the experiments are performed on an 8-core Intel(R) Xeon(R) CPU E3-1240 V2@3.4GHz,3.4GHz CPU, 16GB RAM, NVIDIA GeForce GTX 780Ti graphics card.

**Algorithm 2** Online Testing of Hierarchical 3D Full-body Pose Regression

---
**Require:** depth map $\mathbf{I}$, initial pose $\mathbf{q}^0$ for a testing sample
**Require:** Different stages' regressor $\{R^{s,h}\}_{h=1}^{H_r}, s \in \{r,t,l\}$
**Ensure:** result pose $\widetilde{\mathbf{q}}$
1: **for** $s \in \{r,t,l\}$ **do**  ▷ Traverse all the stages
2:    **for** $h = 1 \to H_s, s = r,t,l$ **do**  ▷ estimate current stage's pose
3:      $\mathbf{q}^{s,h} = \mathbf{q}^{s,h-1} + \widetilde{\beta}^{s,h} \cdot R^{s,h}(\mathbf{I}, \mathbf{q}^{s,h-1})$  ▷ update current stage's pose
4:    **end for**
5:    Initialize $\mathbf{q}^t$ as canonical torso pose on $\mathbf{q}^{s,H_s}$
6:    $\widetilde{\mathbf{q}} = \mathbf{q}^{s,H_s} \cup \mathbf{q}^0$  ▷ update whole body pose
7: **end for**

---

### 5.1 Hyper Parameters Evaluations

We evaluate the influence for the accuracy of pose estimation from different parameters in this section, which can give us the best parameter combination through experiments. These parameters include the depth of trees, the number of tress per stage, the number of stages and the maximum probe offset of the canonical poses feature extraction. All the training samples are randomly taken from the total training sample which consists of 200K images (up to 1.2 hour motion sequence). The remaining 50K samples are used as testing samples (up to 0.3 hour motion sequence). All the experiments keep all the other parameters the same and change only the testing parameters.

**Depth of trees** Theoretically, for the fixed amount of training samples, the depth of the decision tree will directly influence the complexity of the cascaded regressor and in turn influence its generalization ability. Lower depth will lead to lower complexity and lower estimation accuracy of training samples due to under-fitting. Higher depth brings higher estimation accuracy, but lower accuracy of testing samples due to the overfitting. As a result, the tree depth is one of the most important model parameter (another one is the maximum probe offset of the canonical pose). In Figure 5a, we show the comparison results on four different depths (5, 10, 15, 20). For 200K training samples, we can get the best estimation result on the depth of 15, and the RMSE of each joint of 60% testing samples is under 4cm and 80% under 6cm. When the depth of the trees is 10, the RMSE of only 40% testing sample is under 4cm and 70% under 6cm, which is merely better than the case of depth 5. So for the cases of 5 and 10, the cascaded regressors are possibly under-fitting, For depth of 20, which is worse than 15, the RMSE of only 50% testing samples are under 4cm and 80% under 6cm, which means possible overfitting occurred. These experiments shows that with proper depth of trees, we can avoid under-fitting and overfitting problems.

**Number of trees per stage** The number of trees per stage can also have some influences on the accuracy of the 3D human pose estimation result. We evaluate our algorithm on four different tree numbers (1, 8, 16, 24). For 200K training samples, we can get the best estimation result when we use 24 trees per stage. In Figure 5b, for those the RMSE under 4cm and 6cm, the ratio drop to 55% and 85% respectively. When the number of trees decreases to 16, for those the RMSE under 4cm and 6cm are dropping to 35% and 70%. When the number of trees reaches 8, there are only 25% and 55% of training samples whose RMSE are under 4cm and 6cm, respectively, and when we only use 1 tree in each stage, the number drops to 10% and 35%. This experiment shows that, for fixed number of training samples, increasing the number of trees can efficiently increase the accuracy of 3D human pose estimation. Additionally, by incorporating more trees in each stage, we can make the estimated poses smoother. However, the computing efficiency should also be considered, and we employ 16 trees per stage in our experiments for the best balance between accuracy and efficiency.

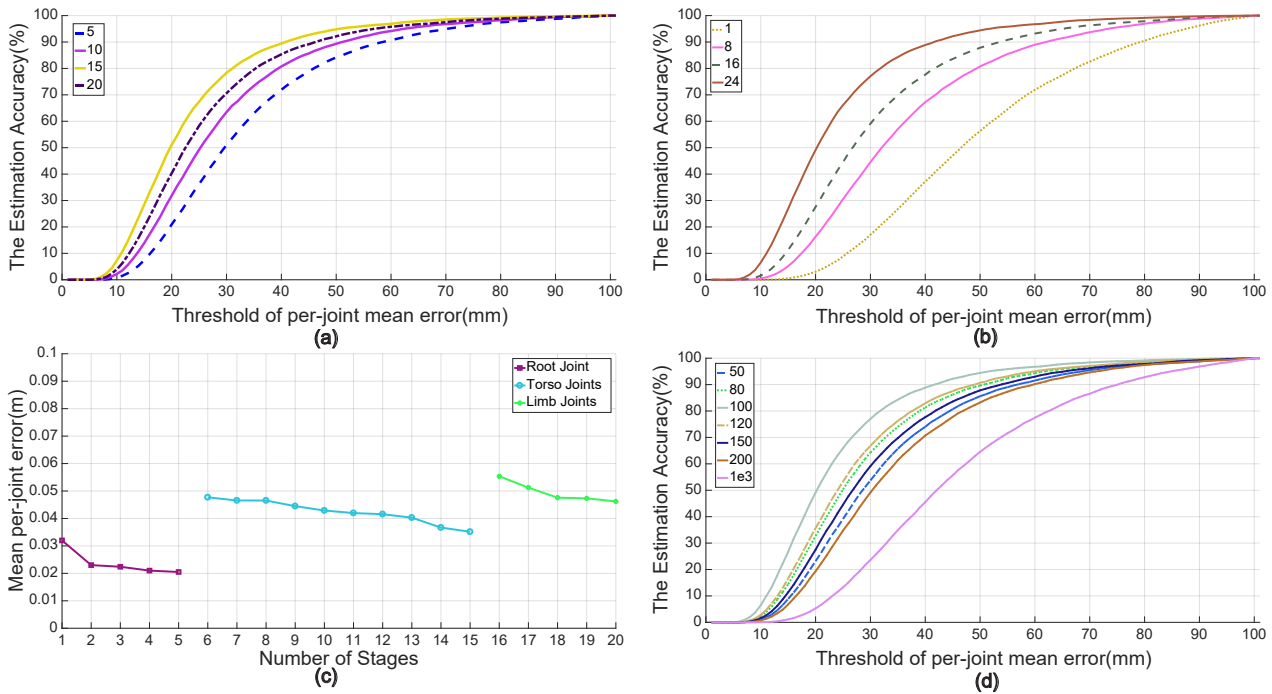

Figure 5: The comparison of different hyper parameters. We define the estimation accuracy as the percentage of per-joint mean error exceed the threshold D (a) The estimation accuracy over different tree depth. (b) The estimation accuracy over different tree numbers. (c) Average joint errors of root, torso and limbs stages. (d) The estimation accuracy over different probe offset.

**Number of stages** We build our regressor based on the hierarchy regression method. We set different stage numbers for the root stage, the torso stage and the limb stage. The Figure 5c shows the estimation error under different number of stages. As we can see from the figure, with the increasing of the number of stages, the curve of the estimation tend to be flat with the number of these three stages separately equal to 5, 10 and 5, which means the current number of stages is already enough. What should be noted is that due to the hierarchy relationship between different stages, we only estimate the error of certain joint in each stage. For example, we only estimate the estimation error of the root joint in the root stage.

**Maximum probe offset** The maximum probe offset is another biggest influence factor on the accuracy of pose estimation. The Figure 5d shows the change of the 3D human pose estimation error with the change of different maximum probe offset. We carry out our experiments on seven different maximum probe offsets, 50mm, 80mm, 100mm, 120mm, 150mm, 200mm and 1m . What should be noted is that the length and radius of root section of the torso section is larger than that of the limbs, so the sampling radius is usually larger than the maximum probe offset. From the error curve of Figure 5d, we can see that setting the maximum probe offset to 100mm, the average per joint error does not exceed 4cm and 6cm for 55% and 85% of the testing samples respectively. Both increasing or decreasing the value of the maximum probe offsets will lead to the decrease of the estimation error, which shows that the extracted feature from unreasonable maximum probe offsets may have weak classification ability.

### 5.2 Regression Objective

We evaluate three different kinds of regression objectives in this section, including the 3D joint position, the rotation angle between joint( Euler angle) and 3D pose changing gradient. According to the experiments, we find that regarding the residual function as the pose changing gradient can achieve more reasonable results than the Euclid distance between the joint angle or the 3D joint position, as shown in Figure 6. Because of the singularity of the Euler angle, especially for those with 3 degrees of freedom such as r/lhumerus and r/lfemur, directly calculating the joint angle may cause the ambiguous rotation and get the unreasonable regression result. Directly doing the regression for the 3D joint position cannot guarantee the bone length. We then adopt the 3D pose changing gradient as the regression objective, which essentially makes a combination of the 3D joint position and the joint angle. It can maintain the bone length while avoiding the ambiguity caused by joint angle.

### 5.3 Accuracy Comparison with Alternatives

We designed comparison experiment for the accuracy of three different 3D pose estimation methods, namely, our method, cascaded method without kinematic model [22], and Kinect V2's method. During the comparison, we adopt six different kinds of motion sequence including walking, running, kicking, jumping, boxing and hand waving (see Figure 7). In Figure 8 , we show the root mean square error (RMSE) of each joint on all testing data from CMU mocap database (The five duplicated 'Site' from left to right are the representations of right hand end site, left hand end site, head end site, right toes end site, and left toes end site, respectively). The average testing accuracy of each method is: 2.44 cm, 5.53 cm and 7.62 cm. Comparing with the random forest based method [21, 22], our cascaded regression method adopts kinematic constraint in the feature extraction from the depth image, and use the gradient residual as the regression task, which can ensure a more reasonable reconstruction result in occlusion. We also evaluated the generality of our method on Human3.6M dataset, and some results of complicated scenario like a walking sequence with significant occlusions (missing arm when turning around) and connected cross(arm and torso) in the image are shown in Figure 9. The testing results verified the generality of our method.

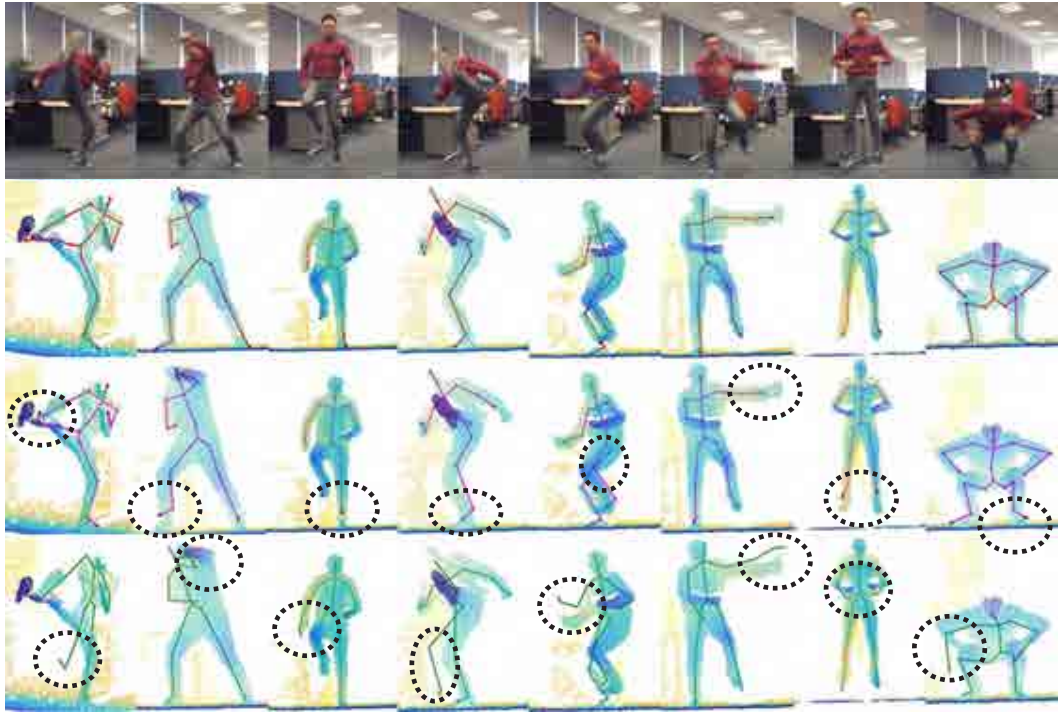

Figure 7: The comparison of 3D human pose estimation accuracy. The RGB image captured by Kinect(First Row); The results with kinematic model(Second Row); The results without kinematic model [22](Third Row); The results of Kinect V2.0(Fourth Row). The false parts are marked with a circle.

## 5.4 Runtime Comparison with Alternatives

Our algorithm only focuses on those pixels near the 3D kinematic joints, while the method of [13, 21] needs to do the estimation for every foreground pixel in the depth image. To achieve the real-time function, we adopt the hardware such as GPU and the parallel computing algorithm. Compared with other method, our method requires less calculation. The comparison of the algorithm efficiency on the same platform is shown in Table 1, which supports our hypothesis that our algorithm has advantages in the computing efficiency and the requiring resources and is totally capable for super real-time application on normal computing hardware. What should be noted is that the cascaded regression method without kinematic model is faster than that with kinematic model due to the reduced calculation amount. However, we concentrate on the accuracy the most, and the sacrificed efficiency is tolerable.

Table 1: The comparison of computing efficiency and resources

| Method Name | FPS | Calculation |
|---|---|---|
| Kinect v2.0 | 30fps | GPU multi-thread |
| With Kinematic Model | 100fps | CPU single thread |
| Without Kinematic Model | 120fps | CPU single thread |

## 6 Conclusion

In this paper, we present a cascaded regression model-based human pose estimation method. The key idea is to adopt a kinematic model during the cascaded regression process. The biggest advantage of our method is that the bone length can be preserved during the whole 3D pose estimation procedure, which leads to more effective features and more accurate. Our method can be used as initialization procedure when combining with tracking methods. The experiments show that with proper hyper-parameters, our method can achieve higher accuracy and faster results compared with previous methods.

There are two main limitations about our approach. One is that our method is not suitable for human pose estimation in outdoor environments due to the natural drawback of current depth camera. The other one is that our method will fail when significant occlusions or self occlusions occurs, such as rolling on the ground. One possible solution to address above two problems is to incorporate RGB images and pose prior knowledge learned from a larger pose database.

### Acknowledgments

This work was supported by the National Natural Science Foundation of China under Grant No.61772499, the Knowledge Innovation Program of the Institute of Computing Technology of the Chinese Academy of Sciences under Grant No.ICT20166040 and the Science and Technology Service Network Initiative of Chinese Academy of Sciences under Grant No.KFJ-STS-ZDTP-017.

### References

[1] Cmu mocap database. http://mocap.cs.cmu.edu.
[2] Kinect 2017, https://developer.microsoft.com/zh-cn/windows/kinect.
[3] A. Baak, M. Muller, G. Bharaj, H. Seidel, and C. Theobalt. A data-driven approach for real-time full body pose reconstruction from a depth camera. pp. 1092–1099, 2011.
[4] L. Breiman. Random forests. Machine learning, 45(1):5–32, 2001.
[5] K. Buys, C. Cagniart, A. Baksheev, T. De Laet, J. De Schutter, and C. Pantofaru. An adaptable system for rgb-d based human body detection and pose estimation. Journal of Visual Communication and Image Representation, 25(1):39–52, 2014.
[6] X. Cao, Y. Wei, F. Wen, and J. Sun. Face alignment by explicit shape regression. International Journal of Computer Vision, pp. 2887–2894, 2012.
[7] D. Comaniciu and P. Meer. Mean shift: A robust approach toward feature space analysis. IEEE Transactions on pattern analysis and machine intelligence, 24(5):603–619, 2002.
[8] M. Ding and G. Fan. Articulated and generalized gaussian kernel correlation for human pose estimation. IEEE Transactions on Image Processing, 25(2):776–789, 2016.
[9] P. Dollar, P. Welinder, and P. Perona. Cascaded pose regression. Computer Vision and Pattern Recognition, pp. 1078–1085, 2010.

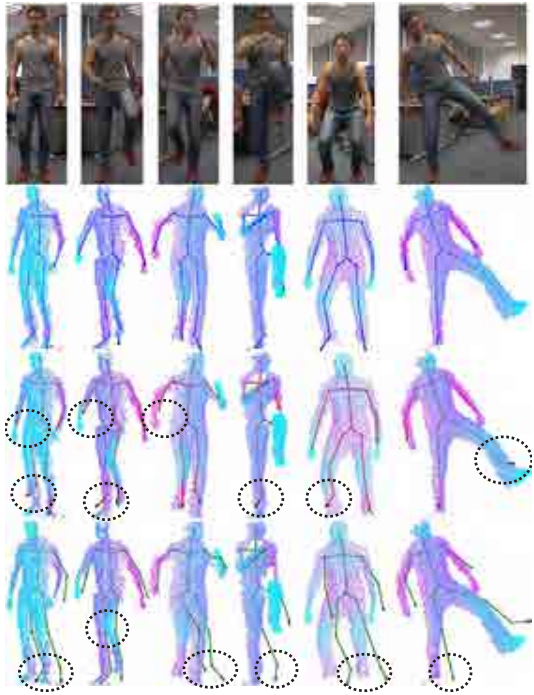

Figure 6: The comparison of different regression objective. The RGB images captured by Kinect(First row); The results of pose gradient(Second row); The results of 3D joint center(Third row); The results of 3D rotation angle(Euler angle, Fourth row); All the results are obtained through cascaded manner, with the same hyper-parameters. The false parts are marked with a circle.

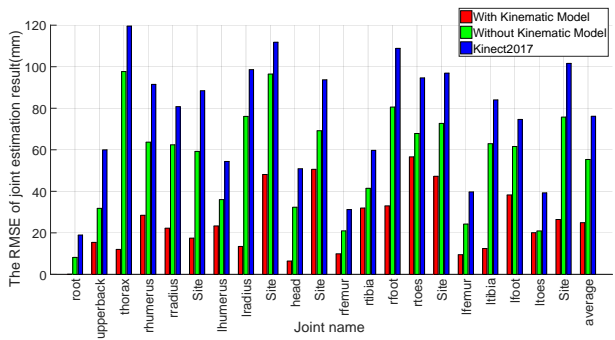

Figure 8: Comparison of different methods on the dataset. The root mean square error of the estimation results over different joints. The RMSE of the root joint with kinematic model is too small to be presented in this figures. The five duplicated 'Site' from left to right are the representations of right hand end site, left hand end site, head end site, right toes end site, and left toes end site, respectively.

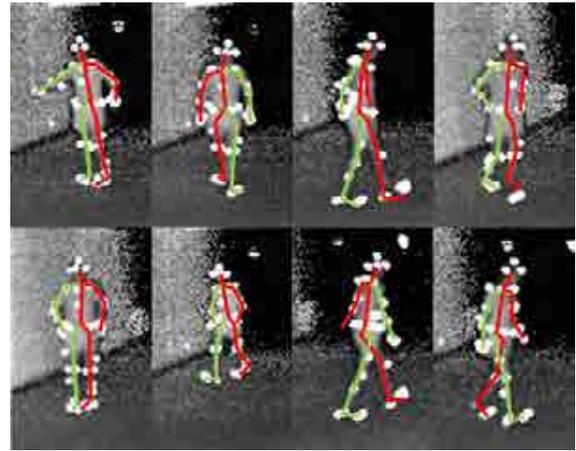

Figure 9: Results of our method on Human3.6M dataset [16]. A walking sequence with significant occlusions (missing arm when turning around) and cross connected (arm and torso) in the image. Red parts: left and torso body parts; Green parts: right body parts.

[10] M. Firman. Rgbd datasets: Past, present and future. computer vision and pattern recognition, pp. 661–673.
[11] J. H. Friedman. Greedy function approximation: a gradient boosting machine. Annals of statistics, pp. 1189–1232, 2001.
[12] J. H. Friedman. Stochastic gradient boosting. Computational Statistics & Data Analysis, 38(4):367–378, 2002.
[13] R. Girshick, J. Shotton, P. Kohli, A. Criminisi, and A. Fitzgibbon. Efficient regression of general-activity human poses from depth images. In 2011 IEEE International Conference on Computer Vision (ICCV),, pp. 415–422. IEEE, 2011.
[14] M. Gleicher. Retargetting motion to new characters. In Proceedings of the 25th annual conference on Computer graphics and interactive techniques, pp. 33–42. ACM, 1998.
[15] S. Handrich and A. Alhamadi. A robust method for human pose estimation based on geodesic distance features. Systems, Man and Cybernetics, pp. 906–911, 2013.
[16] C. Ionescu, D. Papava, V. Olaru, and C. Sminchisescu. Human3.6m: Large scale datacsets and predictive methods for 3d human sensing in natural environments. IEEE Transactions on Pattern Analysis and Machine Intelligence, 36(7):1325–1339, jul 2014.
[17] A. Natekin and A. Knoll. Gradient boosting machines, a tutorial. Frontiers in neurorobotics, 7, 2013.
[18] S. Park, J. Y. Chang, H. Jeong, J.-H. Lee, and J.-Y. Park. Accurate and efficient 3d human pose estimation algorithm using single depth images for pose analysis in golf. In 2017 IEEE Conference on Computer Vision and Pattern Recognition Workshops (CVPRW), pp. 105–113. IEEE, 2017.
[19] T. Probst, A. Fossati, and L. Van Gool. Combining human body shape and pose estimation for robust upper body tracking using a depth sensor. pp. 285–301, 2016.
[20] B. Sapp, A. Toshev, and B. Taskar. Cascaded models for articulated pose estimation. European Conference on Computer Vision, pp. 406–420, 2010.
[21] J. Shotton, T. Sharp, A. A. Kipman, A. Fitzgibbon, M. J. Finocchio, A. Blake, M. Cook, and R. Moore. Real-time human pose recognition in parts from single depth images. Communications of The ACM, 56(1):116–124, 2013.
[22] X. Sun, Y. Wei, S. Liang, X. Tang, and J. Sun. Cascaded hand pose regression. Computer Vision and Pattern Recognition, pp. 824–832, 2015.
[23] W. Swann. A survey of non-linear optimization techniques. FEBS letters, 2:S39–S55.
[24] J. Taylor, J. Shotton, T. Sharp, and A. W. Fitzgibbon. The vitruvian manifold: Inferring dense correspondences for one-shot human pose estimation. Computer Vision and Pattern Recognition, pp. 103–110, 2012.
[25] S. Xia, L. Gao, Y.-K. Lai, M.-Z. Yuan, and J. Chai. A survey on human performance capture and animation. Journal of Computer Science and Technology, 32(3):536–554, May 2017.
[26] H. Yasin, U. Iqbal, B. Kruger, A. Weber, and J. Gall. A dual-source approach for 3d pose estimation from a single image. computer vision and pattern recognition, pp. 4948–4956, 2016.
[27] X. Zhou, X. Sun, W. Zhang, S. Liang, and Y. Wei. Deep kinematic pose regression. European Conference on Computer Vision, pp. 186–201, 2016.